\documentclass[%
 reprint,
superscriptaddress,     % Comment this line if you want groupwise author credits
 amsmath,amssymb,
 aps,
]{revtex4-2}

\usepackage{graphicx}% Include figure files
\usepackage{dcolumn}% Align table columns on decimal point
\usepackage{bm}% bold math
\usepackage[usenames,dvipsnames]{color} %to color texts
\usepackage[normalem]{ulem} % to strike out lines
\usepackage[dvipsnames]{xcolor}

\usepackage{hyperref}% add hypertext capabilities

\begin{document}

\title{Social Phases in Human Movement}

\author{Yi Zhang}
\affiliation{%
Department of Physics, University of Miami, Coral Gables, Florida 33146, USA}%
\author{Debasish Sarker}
\affiliation{%
Department of Physics, University of Miami, Coral Gables, Florida 33146, USA}%
\author{Samantha Mitsven}
\affiliation{%
Department of Psychology, University of Miami, Coral Gables, Florida 33146, USA}%
\author{ Lynn Perry}
\affiliation{%
Department of Psychology, University of Miami, Coral Gables, Florida 33146, USA}%
\author{Daniel Messinger}
\affiliation{%
Department of Psychology, University of Miami, Coral Gables, Florida 33146, USA}%
\author{Chaoming Song}
\email{c.song@miami.edu}
\affiliation{%
Department of Physics, University of Miami, Coral Gables, Florida 33146, USA}%

%\date{\today}
\begin{abstract}
Recent empirical studies found different thermodynamic phases for collective motion in animals. However, such a thermodynamic description for human movement remains unclear,  mainly due to the limited resolution of existing tracking technologies. In this Letter, we used a new ultra-wideband radio frequency identification technology to collect high-resolution spatial-temporal data for classroom movements. We measured the ``effective temperatures" for the underlying systems, and found two social phases, gas and liquid-vapor coexistence, associated with different temperatures. We also proposed a simple statistical physics model that can reproduce different empirically observed phases.
\end{abstract}

\maketitle

\newpage

Collective motion is a ubiquitous phenomenon in many biological and social systems including bacteria \cite{zhang2010collective}, animals \cite{parrish1999complexity}, and humans \cite{vicsek2012collective}. Over the past two decades, substantial research has been devoted to the understanding of human movement, thanks in part to the availability of large-scale human mobility data collected by a broad range of new technologies including banknote circulation \cite{brockmann2006scaling}, mobile-phone records \cite{gonzalez2008understanding,song2010modelling,simini2012universal}, Bluetooth scanning \cite{sekara2016fundamental} and GPS tracking \cite{firth2020using,nagy2010hierarchical}. Initial empirical observations suggest that human movement at the population level satisfies a generalized continuous-time random walk \cite{brockmann2006scaling}, while later work shows that individuals’ mobility patterns are better captured by a more sophisticated exploration and preferential return mechanism \cite{song2010modelling}. However, these empirical studies focus mainly on the asymptotic behavior of moving individuals; little emphasis is placed on social interaction among them.

%Previous paragraph: In parallel, recent empirical studies on animals use synchronized cameras to reconstruct high-resolution moving trajectories for swarming insects \cite{sinhuber2017phase,attanasi2014collective}, fish \cite{katz2011inferring,herbert2011inferring}, and birds \cite{bialek2012statistical}. These researches find coexistence of a dilute ”vapor” and a ”condensed” phase in equilibrium in insect swarm groups due to group interactions, implying that the interactions among individuals are not negligible. Indeed, early research on pedestrian crowds \cite{vicsek1995novel, helbing1995social,helbing2000simulating}, and vehicular traffic \cite{helbing2001traffic} suggested that group interactions have a large impact on  human movements. However, these studies focus mainly on a macroscopic picture of human movement patterns, and it is difficult  to investigate individual interactions empirically at the microscopic level.

In parallel, theoretical work on the social force model \cite{vicsek1995novel,helbing1995social,helbing2000simulating,helbing2001traffic} suggested a potential liquid-vapor transition in human movements because of social interaction \cite{vicsek1995novel, gregoire2004onset}. Moreover, recent empirical studies of animals use synchronized cameras to reconstruct high-resolution moving trajectories for swarming insects \cite{sinhuber2017phase,attanasi2014collective}, fish \cite{katz2011inferring,herbert2011inferring}, and birds \cite{bialek2012statistical}. This research indicates the coexistence of a ``condensed" and a dilute ``vapor" phase in equilibrium in insect swarms, calling attention to a thermodynamic description of individual movements. However, to our best knowledge, an empirical exploration of such macroscopic phases in human movement is missing in the current literature, partly due to a lack of high-resolution records of individuals’ trajectories during social events.

%Old: To investigate the impact of social interactions on human movements, one requires a high-resolution record of moving trajectories. Existing approaches, usually by means of GPS [cite] and mobile phone data [8], are incapable of such a task because the social interactions often occur within 1-2 meters which is beyond the spatial resolution of these traditional techniques. In addition, social events typically take place indoors where GPS and cell phone tracking are largely limited to track social contacts. The badge-based technologies  \cite{messinger2019continuous,veiga2017social} allows to detect social contacts when two individuals are within a certain proximity, however, these can neither give an explicit interpretation of the defined social distance, nor capture the exact locations where the interactions take place.

Existing tracking techniques such as GPS \cite{firth2020using,nagy2010hierarchical} and mobile phone data \cite{gonzalez2008understanding,song2010modelling} are too coarse for such a task because social interactions often occur within 1-2 meters \cite{messinger2019continuous}, beyond the spatial resolution of these traditional techniques. In addition, social events may take place indoors where GPS and cell phone tracking are largely limited to track social contacts. Badge-based technologies \cite{sinhuber2017phase,attanasi2014collective,sekara2016fundamental,stehle2011high} and Bluetooth scanning \cite{sekara2016fundamental} allow detection of  social contacts when two individuals are within a certain distance. However, badge-based technologies neither give an explicit description of the defined social distance nor capture the exact locations where interactions take place.

%Old: In this letter, we investigate the human movement driven by social interactions within the classroom using a commercial tracking system based on ultra-wideband radio frequency identification (UWB-RFID). The UWB-RFID allows us to measure the locations and orientations of each individual with a high spatial-temporal resolution. This allows us to measure a number of physical quantities  such as mean-square displacement, radial distribution and correlations function. Assuming the validity of the fluctuation-dissipation theorem, we next estimate the “effective temperature” of three different classrooms. We find two different “social” phases at different temperatures, in analogy to fluid and liquid for molecules: (i) \textit{gas-like phase} where individuals move freely without forming small social groups, and (ii) \textit{liquid-vapor-coexistence-like phase} where individuals form small social groups with free-moving ones joining and leaving. We find that the “effective temperature” for the social gas-like phase is significantly higher than the one for liquid-vapor coexistence phase. Finally, we propose a simple model that integrates the interplay between separation distances and orientations. We perform numerical simulations of the proposed model, and find it is capable of reconstructing the empirically observed social phases. Moreover, we show that physical quantities measured in our model agree qualitatively with the experimental observations. 

In this Letter, we investigate human movement in classrooms using a commercial tracking system based on ultra-wideband radio frequency identification (UWB-RFID). The UWB-RFID allows us to measure the locations and orientations of each individual with a high spatial-temporal resolution, and estimate the ``effective temperature" of classroom movement. By measuring local density distribution, we discovered two different ``social" phases at different temperatures: (i) gas-like phase where individuals move freely without forming small social groups, and (ii) liquid-vapor coexistence-like phase where individuals form small social groups with free-moving individuals joining and leaving these groups. We find that the effective temperature for the gas-like social phase is significantly higher than the effective temperature of the liquid-vapor coexistence phase. Finally, we propose a new social force model inspired by the empirical data. Numerical simulations of our model successfully reproduce the empirically observed social phases. Finally, we show that physical quantities measured in our model agree qualitatively with the experimental observations.

%Old: {\it Data Collection.} Using UWB-RFID technology, we collect human movement datasets in classrooms of three schools. Receivers in the corners of each classroom received signals at $2\sim4 Hz$ allowing continuous measurement of individual’s locations with a $13-cm$ spatial accuracy \cite{matakos2017measuring,irvin2018automated}. D1) comprises records in $2018-2019$ academic year on the classroom movements of two preschools in the US, each consisting of $10$ (D1A) and $17$ (D1B) children aged between $2.5\sim4$ years, respectively. A total of $13$ three-hour observations in class D1A and $16$ observations in class D1B were recorded. Vests with RFID tags over children’s hips were specially designed to be worn by them during their regular classroom activities. Unifying the spatial location information from tags on the left and right side of each vest, orientation information of each child was also extracted. D2) comprises records in  children aged around $5$ years in one kindergarten in Germany in the fall $2014$. Children were equipped with a single tag on their wrists,  which allowed the UWB-RFID to trace their locations only. Three observations were recorded during a $1$-hour free-play period.

{\it Data Collection.} Using UWB-RFID technology, we collected human movement datasets in classrooms in three schools. Receivers in the corners of each classroom received UWB-RFID signals at $2\sim4$~Hz, allowing continuous measurement of individual’s locations with a $13$~cm spatial accuracy \cite{matakos2017measuring, irvin2018automated}. D1) comprises records in $2018-2019$ academic year on the classroom movements of two preschools in the US, each consisting of $10$ (D1A) and $17$ (D1B) children aged between $2.5\sim4$ years, respectively. A total of $13$ three-hour observations in class D1A and $16$ observations in class D1B were recorded. Vests with RFID tags over children’s hips were specially designed to be worn by them during their regular classroom activities. Unifying the spatial location information from tags on the left and right side of each vest, the orientation information of each child was also extracted. D2) comprises records of children aged around $5$ years in one kindergarten in Germany in the fall $2014$. Children were equipped with a single tag on their wrists,  which allowed the UWB-RFID to trace their locations only. Three observations were recorded in D2, each involving a one-hour free-play period (see SM Section A for details).

%Old: {\it Empirical measures.} A preliminary visualization (Fig. 1a) of nine individuals from class D1 during a continuous class time period shows that the participants are not uniformly located in the whole classroom space, implying that they may form small social clusters groups. Although individuals are confined in these clusters by means of social interaction, they become more mobile once they are out of the social groups. Such social clusters are dynamically stable with individuals joining and leaving, exhibiting a similar pattern as liquid droplets. Yet, there also exist free-moving individuals with relatively higher mobility, displaying a similar pattern as gas. For instance, the individual represented by the red arrow is first within the social group near the left boundary, then this individual moves close to the right corner interacting with the purple individual, and eventually, approaches the green individual. Based on visual inspection of arrow density we find that individuals spend more time in social groups than staying lonely by themselves. Furthermore, individuals’ orientation represented by the arrow are shown to be correlated within  groups. These observations suggest that the dynamics of such systems is driven by social interaction. 

{\it Empirical measures.} A preliminary visualization (Fig. 1a) of nine individuals from class D1 during a continuous class time period shows that the participants are not uniformly located in the classroom space, suggesting they may form small social groups. The emergence of such groups is consistent over different observations with individuals joining and leaving the groups. It appears that the density of children varies both by physical location and over time. Furthermore, individuals' orientations, represented by their arrows, appear to be correlated within groups.

\begin{figure}
  \includegraphics[width=1\linewidth]{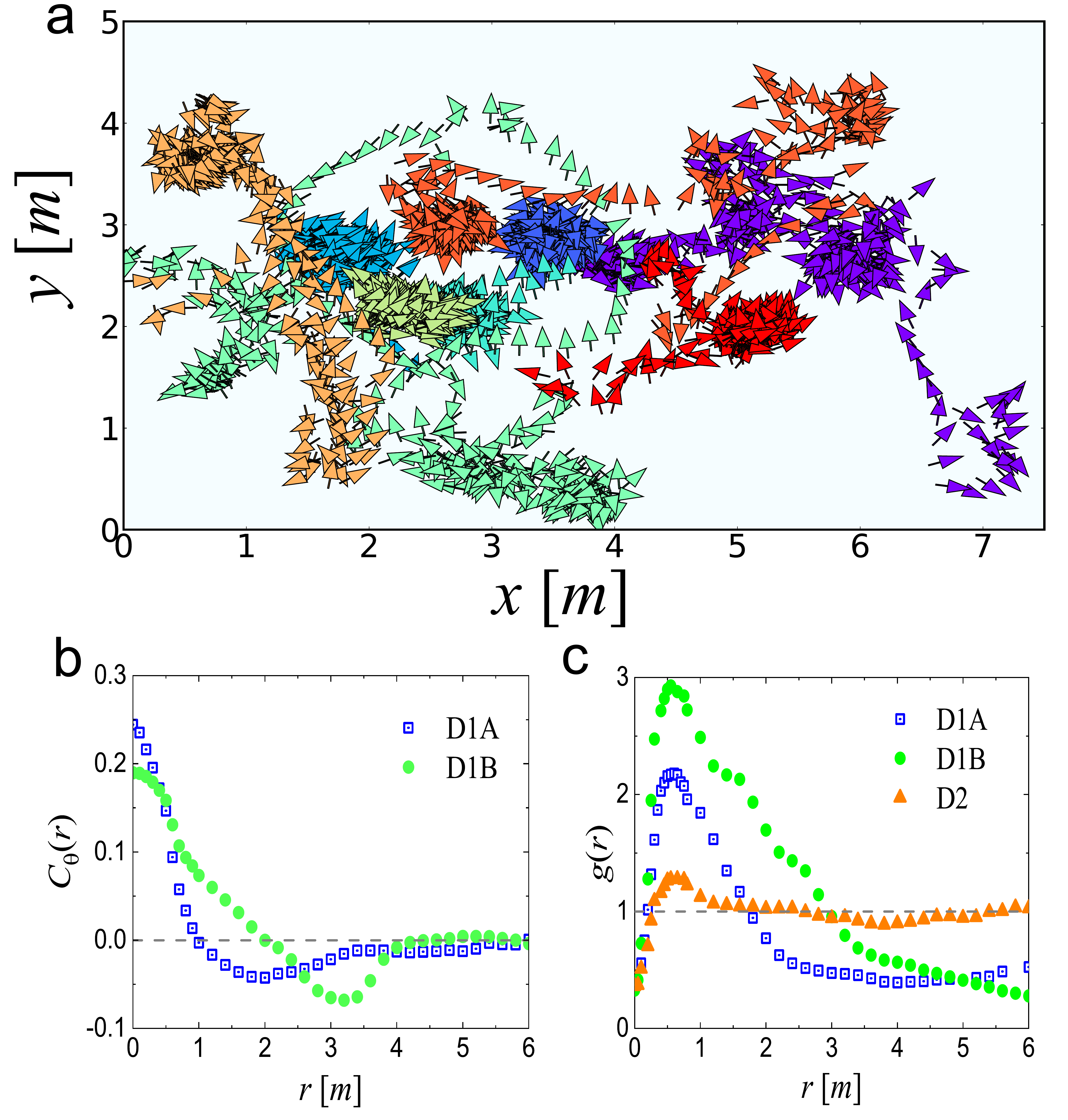}
  \caption{\textbf{Trajectories and distributions.} (a) An illustration of 300 seconds of the trajectories of nine individuals from the D1A, colors and arrows identify the specific individuals and their orientations, respectively. (b) Angular correlation function $C_{\theta}(r)$ for the dataset D1. $C_{\theta}(r) > 0$ and $<0$ correspond to ferromagnetic and antiferromagnetic orientations, respectively. The Grey dashed line represents no correlation. \textit{Note dataset D2 has no orientation measurement}. (c) Radial distribution function $g(r)$ for all three datasets D1-2. 
}
  \label{fig:cluster}
\end{figure}

%Old: To quantify orientational interaction among pairs, we compute the angular correlation function \cite{fanstudy,nho2002spin} $C_{\theta}(r)=\langle\cos{(\theta_i(0)-\theta_j(r))}\rangle$ that measures the alignment of pair angles as a function of  separation distance $r$, averaged over all pairs. For instance, $C_{\theta}(r)=1$ indicates a parallel alignment, i.e., $\theta_i=\theta_j$ whereas $C_{\theta}(r)=-1$ represents an opposite alignment as $\theta_i=-\theta_j$. Figure 1b plots $C_{\theta}(r)$ for datasets D1A and D1B, showing that $C_{\theta}(r)$ is positive when $r$ is small and becomes negative at $r\approx2-4m$. When \emph{r} becomes large, $C_{\theta}(r)$ ceases as the correlation declines. This observation suggests at short distances individuals are more likely to be oriented parallel to each other like ferromagnets due to the domination of side-by-side interactions. On the other hand at longer distances they tend to be oppositely oriented like anti-ferromagnets due to the contribution of face-to-face interaction.

To quantify orientational interaction among pairs, we compute the angular correlation function \cite{nho2002spin} $C_{\theta}(r)=\langle\cos{(\theta_i(0)-\theta_j(r))}\rangle$ that measures the alignment of pair angles as a function of  separation distance $r$, averaged over all pairs. For instance, $C_{\theta}(r)=1$ indicates a parallel alignment (side-by-side), i.e., $\theta_i=\theta_j$ whereas $C_{\theta}(r)=-1$ represents an opposite (face-to-face) alignment as $\theta_i=-\theta_j$. Figure 1b plots $C_{\theta}(r)$ for datasets D1A and D1B, showing that $C_{\theta}(r)$ is positive when $r$ is small and becomes negative at $r\approx2\sim4m$. When \emph{r} becomes large, $C_{\theta}(r)$ approaches zero as the correlation declines. This observation suggests that at short distances individuals are more likely to be oriented parallel to each other, like ferromagnets, due to the domination of side-by-side interactions. On the other hand, at longer distances they tend to be oppositely oriented like anti-ferromagnets due to the contribution of face-to-face interaction.

%Old: We next measure the radial distribution function (RDF), $g(r)$, the probability of finding a pair of individuals at a separation distance $r$, relative to random movements. $g(r) = 1$ indicates no pair correlations, where $g(r)>1$ and $g(r)<1$ correspond to attractive and repulsive pair interactions respectively. Figure 1c plots RDF for all datasets, finding similar patterns across different datasets. First, $g(r)$ consistently peaks around $r_0\approx0.8m$, suggesting a typical social contacting distance regardless of experimental settings. Second, individuals are unlikely to appear at very short distances around $r < 0.3m$, due to the physical constraints of their body size. Third, social interaction is persistent over a broad range of distances from $r = 0.3m$ up to $2$--$3m$. On the other hand, the peak values of $g(r_0)$ vary significantly for D1A-D1B, indicating variation in social interaction strength. Besides the weakest interactions observed in D2, we also find its RDF converges to unitary at long distances, a typical pattern for gas. In contrast, we find $g(r)$ is lower than $1$ one asymptotically for D1 datasets, a phenomenon often observed in the liquid-vapor coexistence phase \cite{sinhuber2017phase,doye1996effect,mochizuki2015solid}, suggesting that individuals form social groups in these datasets. This finding implies $D1$ shows a different macroscopic phase compared to the dataset D2, evidenced by social clustering due to a stronger interaction \cite{attanasi2014collective,attanasi2014finite}.

We next measure the radial distribution function (RDF), $g(r)$, the probability of finding a pair of individuals at a separation distance $r$, relative to random movements. $g(r) = 1$ indicates no pair correlations, where $g(r)>1$ and $g(r)<1$ correspond to attractive and repulsive pair interactions respectively. Figure 1c plots RDF for all datasets, finding similar patterns across different datasets. First, $g(r)$ is consistently peaked around $r_0\approx0.8m$, suggesting a typical social contact distance regardless of experimental settings. Second, individuals are unlikely to appear at very short distances around $r < 0.3m$, due to physical constraints of body size. Third, social interaction is persistent over a broad range of distances from $r = 0.3m$ through $2-3$$m$. On the other hand, the peak values of $g(r)$ vary significantly for D1A-D1B, indicating variation in social interaction strength. Not only are interactions weakest in D2, but its RDF also converges to unity at long distances, a typical pattern for gas. In contrast, we find $g(r)$ is lower than $1$  asymptotically for D1 datasets, a phenomenon often observed in the liquid-vapor coexistence phase \cite{sinhuber2017phase,doye1996effect,mochizuki2015solid}. This finding implies that D1 has a different macroscopic phase compared to the dataset D2, evidenced by social clustering due to  stronger interaction \cite{attanasi2014collective,attanasi2014finite}.

\begin{figure}
\includegraphics[width=1\linewidth]{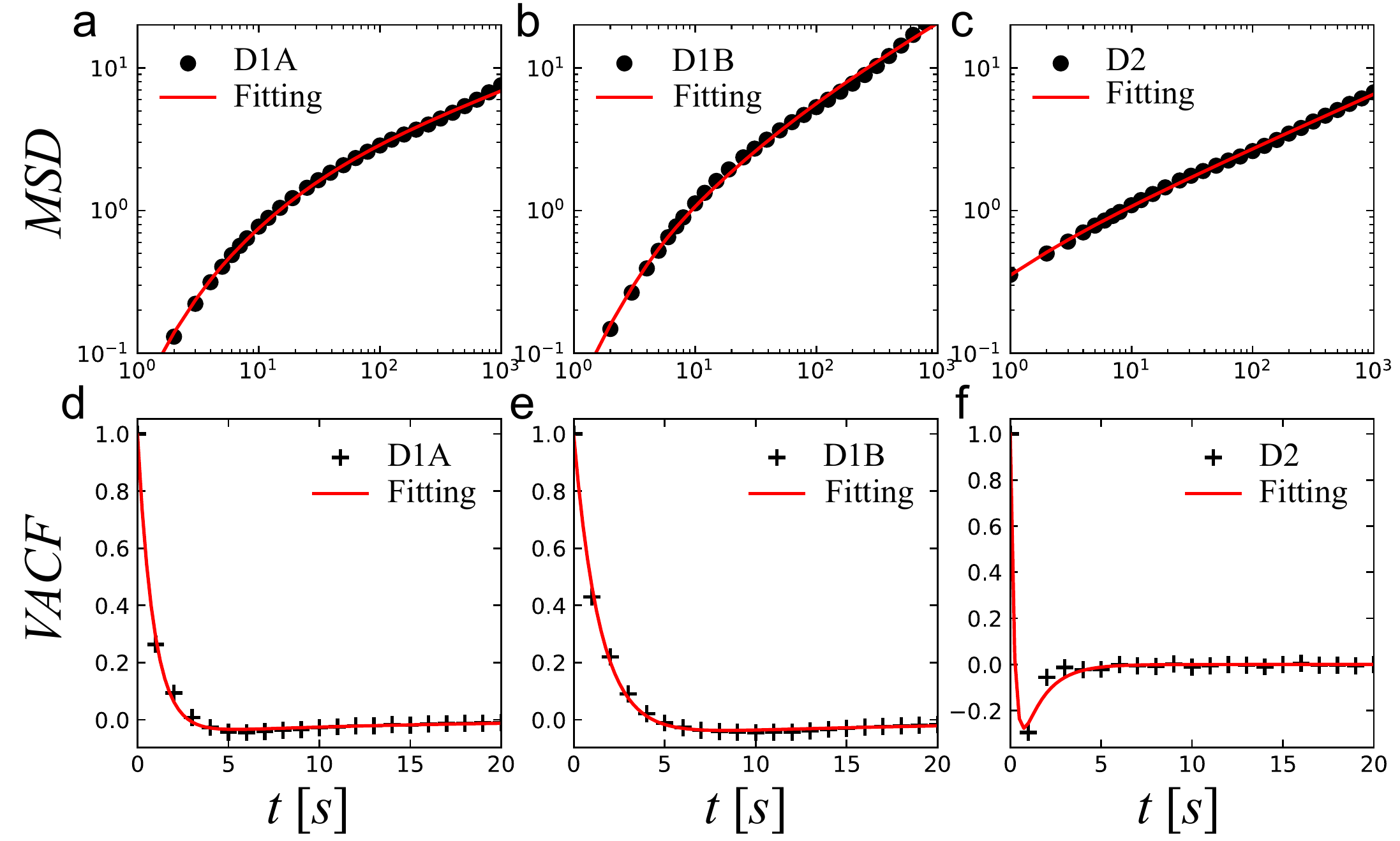}% Here is how to import EPS art
\caption{\label{fig:msd-vacf}\textbf{Empirical dynamical measures.} (a)-(c) Mean square displacement $\langle\Delta r^2\rangle$. (d)-(f) Velocity autocorrelation function $C_v(t)$. Dots represent experimental measures, whereas red curves represent the least-squaring fitting of the theory (see SM Section B).}
\end{figure}

%Old: {\it Effective temperature.} To distinguish between different social phases, we implement tools from statistical physics to explore a central thermodynamic quantity, temperature $T$. The Stokes-Einstein relation $k_BT=D\zeta$ relates temperature in Brownian motion to the diffusion constant $D$ and viscosity $\zeta$. For interactive systems, Stokes- Einstein relation has been generalized as the fluctuation-dissipation theorem  \cite{kubo1966fluctuation,bruna2017diffusion}. To find diffusion constant $D$, we first measure the mean square displacement (MSD), $\langle\Delta r^2\rangle$, that quantifies the variance of individual movements over time t. For two-dimensional Brownian motion the MSD scales linearly with time as $\langle\Delta r^2\rangle = 4Dt$, where the slope is proportional to the diffusion constant $D$. Figure 2a-c plot averaged MSD for datasets D1-D2, showing a linear dependency at a short time scale $t<<\tau_s$, where $\tau_s$ represents a typical time scale for social interaction. We find that $\tau_s\approx3s$ for datasets D1A-D1B, whereas the MSD in D2 does not show a normal diffusion regime, due to experimental limitation of temporal resolution. For a long time $t\gg\tau_s$, individuals diffuse sub-linearly, as $\langle\Delta r^2\rangle\sim t^{\alpha}$, where the exponent $\alpha\approx0.3\sim0.5$ (see Table~\ref{tab:parameters}) indicates a slowdown in movement due to social interaction. We measure the diffusion constant $D$ for D1-D2 as shown in Table I, finding the D2 dataset has the largest diffusion constant.

{\it Effective temperature.} To distinguish between different social phases, we implement tools from statistical physics to explore a central thermodynamic quantity, temperature $T$. The Stokes-Einstein relation $k_BT=D\zeta$ relates temperature in Brownian motion to the diffusion constant $D$ and viscosity $\zeta$. For interactive systems, the Stokes-Einstein relation has been generalized as the fluctuation-dissipation theorem \cite{kubo1966fluctuation,bruna2017diffusion}. To find diffusion constant $D$, we first measure the mean square displacement (MSD), $\langle\Delta r^2\rangle$, that quantifies the variance of individual movements over time $t$. For two-dimensional Brownian motion, the MSD scales linearly with time as $\langle\Delta r^2\rangle = 4Dt$, where the slope is proportional to the diffusion constant $D$. Figure 2a-c plots averaged MSD for datasets D1-D2, showing a linear dependency at a short time scale $t\ll\tau_s$, where $\tau_s$ represents a typical time scale for social interaction. We find that $\tau_s\approx3s$ for datasets D1A-D1B, whereas the MSD in D2 does not show this typical diffusion regime, due to experimental limitation of temporal resolution (see SM Section B2). In the continuation of $\langle\Delta r^2\rangle$, $t\gg\tau_s$; that is, individuals diffuse sub-linearly, as $\langle\Delta r^2\rangle\sim t^{\alpha}$. The exponent $\alpha\approx0.3\sim0.5$ (see SM Section B2) indicates a slowdown in movement due to social interaction. We measure the diffusion constant $D$ for D1-D2 (see SM Section B2), finding that the D2 dataset has the largest diffusion constant.

%Old: In parallel, we also measure the velocity autocorrelation function (VACF), $C_V(t)=\langle\vec{v}(t)\cdot\vec{v}(0)\rangle$  that quantifies the temporal velocity correlation over a time interval t. For Brownian particles, VACF satisfies an exponential decay \cite{kubo1966fluctuation,leegwater1991velocity,li2013brownian}, $C_V(t)=e^{-t/\tau}$, where $\tau=m/\zeta$ represents the typical time scale \cite{kubo1966fluctuation}. We measure VACF for all datasets as shown in Fig. 2d-f. At short times, the VACF follows exponential decay as Brownian particles. However, VACF becomes negative at a longer time, implying attractive social interactions among individuals. We find the empirical VACF can be fitted by $f(t)=Ae^{-t/\tau}/-Be^{-t/\tau_s}$, where former and later terms capture the thermal diffusion and social interaction, respectively. Table~\ref{tab:parameters} reports $\tau$ and $\tau_s$ values for D1-3, showing $\tau\approx0.2\sim1.5s$ and $\tau_s\approx1.2\sim3.5s$. 

In parallel, we also measure the velocity autocorrelation function (VACF), $C_v(t)=\langle\vec{v}(t)\cdot\vec{v}(0)\rangle$ that quantifies the temporal velocity correlation over a time interval $t$. For Brownian particles, VACF satisfies an exponential decay \cite{kubo1966fluctuation,leegwater1991velocity,li2013brownian}, $C_v(t)=e^{-t/\tau}$, where $\tau=m/\zeta$ represents the typical time scale \cite{kubo1966fluctuation}. Figures 2d-f show VACF measured for all datasets.  At short times, the VACF follows exponential decay as Brownian particles. However, VACF becomes negative for a longer time, implying attractive social interactions among individuals. We find the empirical VACF can be fit by $Ae^{-t/\tau}-Be^{-t/\tau'}$, where former and later terms capture the thermal diffusion and social interaction, respectively. Note that $\tau'$ diffs from $\tau_s$ measured in MSD but they are positively correlated (see SM Section B1). 

%Old: The Stokes-Einstein relation allows us to define an effective temperature $k_BT\equiv mD/\tau$. Although such an effective temperature differs from the thermodynamic ones defined in physics, it quantifies the movement activity of classrooms. Table I computes the normalized effective temperature, $\tilde{T}=k_BT/m$ for D1-D2, finding that the effective temperature $\tilde{T}=0.06J/kg$ and $0.10 J/kg$ for D1A and D1B, respectively, while D2 has the highest effective temperature $\tilde{T}=1.47J/kg$, reflecting the fact that D2 is free-play kindergarten. This is also consistent with the $g(r)$ measurement in Fig. 1c which shows the school D2 has the weakest social interaction.

The Stokes-Einstein relation allows us to define an effective temperature $k_BT\equiv mD/\tau$ for human movement. Although such an effective temperature differs from the thermodynamic temperature of physical substances, it quantifies the movement activity of classrooms. We measured the reduced effective temperature, $\tilde{T}\equiv k_BT/m$ for both datasets, finding $\tilde{T} = 0.06J/kg$ and $0.10 J/kg$ for D1A and D1B, respectively, while D2 has the highest effective temperature $\tilde{T}=1.47J/kg$, reflecting the fact that D2 involves only free-play activities (see SM Table S2 for a summary). This is also consistent with the RDF measurement in Figure 1c, showing that dataset D2 has the weakest social interaction.

{\it Local density distribution.} The RDF function suggests that the higher temperature dataset D2 falls into a gas-like social phase. On the other hand, the lower temperature datasets D1A-B appear to be in the liquid-vapor coexistence-like social phase. To distinguish these two different social phases, we calculated the local density distribution (LDD) \cite{hansen1988theory}, $p(\rho)$, that measures the spatial variation of the local number density of particles (see SM Section B3). In the coexistence phase, $p(\rho)$ is bipeaked \cite{gregoire2004onset} because  the particles are inhomogeneously distributed in space where the low and high-density peaks represent vapor and liquid phases, respectively. In contrast, $p(\rho)$ is single-peaked around the average density, $\bar{\rho}$, for the gas phase. Figure 3a-c, plot $p(\rho)$ for datasets D1-D2, showing that $p(\rho)$ is mono-peaked for the high-temperature dataset (D2). In contrast, we find that $p(\rho)$ in D1 agrees well with a bi-gamma distribution, with a vapor density peak around $\rho_v=0.2\bar{\rho}\sim0.4\bar{\rho}$, and a liquid density peak around $\rho_l=1.05\bar{\rho}\sim1.20\bar{\rho}$. This result suggests a liquid-vapor coexistence phase in D1 and a gas phase in D2. While existing models suggest a similar liquid-vapor phase transition \cite{gregoire2004onset}, these transitions are often coupled with a change of orientational order. We measured the magnetization order parameter $m_0 \equiv\frac{1}{N}\left|\left<\left (\sum_{i=1}^{N}\cos\theta_i,\sum_{i=1}^{N}\sin\theta_i \right ) \right> \right|$, for the dataset D1, finding $m_0 \approx 0.001$ and $0.002$ for D1A and D1B respectively. This finding suggests that unlike the existing models \cite{vicsek1995novel,gregoire2004onset}  the liquid-vapor phase transition does not accompany by orientational order changes, partly due to the mixture of ferro- and antiferromagnetic orders at different distances. 

%Old: {\it Phase diagram.} We illustrate a typical $\rho-T$ phase diagram for a single component molecular system in figure 3d. For temperature below the critical temperature $T_c$, the system undergoes a vapor-liquid phase transition where different phases are governed by the system density. The low density regime ($\rho<\rho_v$) corresponds to a vapor phase where the high density regime ($\rho>\rho_l$) represents a liquid. For the density between $\rho_v$ and $\rho_l$, vapor and liquid coexist. The density difference $\Delta\rho=\rho_l-\rho_v\sim(T_c-T)^{\beta}$ decreases as temperature $T$ approaches $T_c$, where the critical exponent $\beta=1/2$ for mean field theory \cite{fisher1964correlation} and $\beta\sim0.35\pm0.15$ experimentally \cite{goodwin1970estimation,hansen1988theory}. The distinction between vapor and liquid disappears when $T = T_c$, and above $T_c$, the system is gas.

{\it Phase diagram.} We illustrate the $\rho-T$ phase diagram for classroom movement in Figure 3d. For temperature below the critical temperature $T_c$, the system undergoes a vapor-liquid phase transition where different phases are governed by the system density. The low-density regime ($\rho<\rho_v$) corresponds to a vapor phase, whereas the high-density regime ($\rho>\rho_l$) represents a liquid. When density is between $\rho_v$ and $\rho_l$, vapor and liquid coexist. The density difference $\Delta\rho=\rho_l-\rho_v\sim(T_c-T)^{\beta}$ decreases as temperature $T$ approaches $T_c$, where the critical exponent $\beta=1/2$ for mean field theory \cite{fisher1964correlation} and $\beta\approx0.35\pm0.15$ experimentally for molecules \cite{goodwin1970estimation,hansen1988theory}. The distinction between vapor and liquid disappears when $T = T_c$, and above $T_c$ the system is gas-like.

%Old: While the social system is different from molecules in many aspects, a similar paradigm seems also to apply to our datasets. In particular, the bi-peaked local density distributions in school D1 suggests they belong to the vapor-liquid coexistence phase. Individuals are less mobile with a relative low temperature. Most individuals are trapped within small social groups, in analogy to droplets for molecular systems. Occasionally, some active individuals escape from social groups and move like a vapor molecule. Social groups are changing dynamically, with people joining and leaving. To quantify the liquid-vapor phase transition, we find $\Delta\rho_{D1}\sim0.9\pm$ for school D1, and $\Delta\rho_{D2}\sim0.6\pm$ for school D2. Since our dataset is limited to two data points in the coexistence phase, we can only roughly estimate the critical temperature $\tilde{T}_c\approx0.11\sim0.12 J/kg$ and $\beta\approx0.3\sim0.6$. In contrast, school D2 falls into the gas phase where individuals move freely without forming social groups under a high temperature environment, in line with its mono-peaked local density distribution. 

While social systems clearly differ from molecules, similar equations seem to apply to our datasets. In particular, the bipeaked local density distributions in school D1 suggest they exhibit a vapor-liquid coexistence phase. Individuals in D1 are less mobile with a relatively low temperature. Most individuals are trapped within small social groups, in analogy to droplets in molecular systems. Occasionally, some active individuals escape from social groups and move like a gas molecule. Social groups evolve dynamically with individuals joining and leaving the group. To quantify the liquid-vapor phase transition, we find $\Delta\rho_{D1}\approx0.9$ for school D1A, and $\Delta\rho_{D2}\approx0.6$ for school D1B. Since our dataset is limited to two data points in the coexistence phase, we can only roughly estimate the critical temperature $\tilde{T}_c\approx0.11\sim0.12 J/kg$ and $\beta\approx0.3\sim0.6$. In contrast, school D2 falls into the gas phase, where individuals move freely without forming social groups under a high-temperature environment, in line with its mono-peaked local density distribution.

\begin{figure}
  \includegraphics[width=1\linewidth]{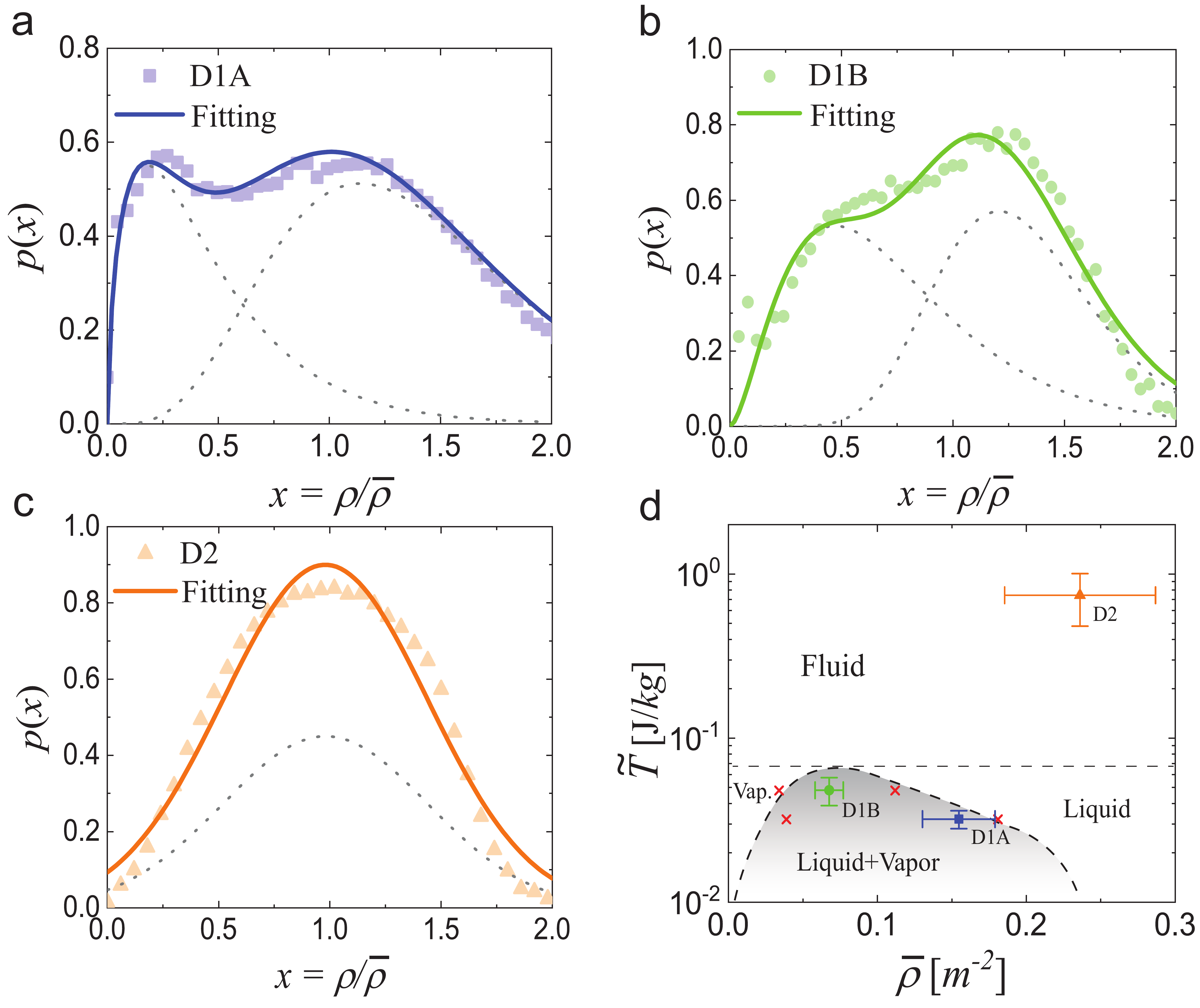}
  \caption{\textbf{Local density distribution measures of empirical data and phase diagram.} (a)-(c) Local density distribution for datasets D1A-B, respectively. $x$ axis is the local density normalized by average density $\bar{\rho}$. Scatter plots represent the empirical measures, and colored curves indicate curve fitting. D1 shows a typical bipeak pattern, implying a vapor-liquid coexistence phase. D2 demonstrates a typical mono-peaked local density distribution, representing a single vapor phase. (d) Density-temperature phase diagram. The grey dashed curves separate the different phase domains. The red crosses are estimated by the local density distributions of D1.}
  \label{fig:localdensitydata}
\end{figure}

%Old: {\it Minimal model.} We proposed a simple model that captures the main features of empirically observed movements. We model individuals as small magnets similar to the classical XY model ~\cite{loft1987numerical}
%\begin{align}
%    U(\vec{r},\theta)=\sum_{i<j}J(r_{ij})\cos{(\theta_i-\theta_j)} \label{eq:xy_model}
%\end{align}
%Unlike the lattice models, our model allows individuals to move continuously in space, driven by the pair interactions based on Eq. (1).  In addition, we also applied soft sphere potential for each individual to model their physical size. We use the Morse potential for the coupling function J(r) which captures the empirical observation for the ferromagnetic interaction at short extent whereas an antiferromagnetic interaction at long distances (Fig. 1b).

%We applied Monte Carlo simulations with $N = 2,500$ individuals for our model with different temperature and density values. For densities comparable to empirical ones, we found  mono-peaked local density distributions for high-temperature values corresponding to a gas phase (Fig. 4a),  whereas bi-peaked distributions for lower temperatures corresponding to a liquid-vapor coexistence phase (Fig. 4b). The insets of Figure 4a,b demonstrate snapshots of movement patterns for these two phases, suggesting that our model reproduces social phases observed empirically.

{\it Simple model.} Unlike existing models \cite{vicsek1995novel, gregoire2004onset}, our empirical observation suggests that liquid-vapor phase transition is independent of the change of orientational order. Inspired by the empirical observation, we model moving individuals as small magnets with
a $XY$-model like interaction \cite{loft1987numerical}, 
\begin{align}
    U(\textbf{r},\theta)=\sum_{i<j}J(r_{ij})\cos{(\theta_i-\theta_j)},
    \label{eq:model}
\end{align}
where $r_{ij} \equiv |\textbf{r}_i-\textbf{r}_j|$, and $J(r)$ is the coupling function that varies by distance. When $J(r)$ is the Heaviside step function, Eq.~(\ref{eq:model}) leads to a social force similar to the Vicsek model \cite{vicsek1995novel}. However, our empirical observation in Figure 1b suggests a more subtle interaction with mixed ferromagnetic and antiferromagnetic effects. Therefore, we use the Morse potential \cite{kaplan2003handbook} for the coupling function $J(r)$, which captures the empirical observation of ferromagnetic interaction at a short extent and an antiferromagnetic interaction at longer distances (Fig. 1b). In addition, we also used a soft sphere potential for each individual to take  their physical size into consideration (see SM Section C).

Equation~(\ref{eq:model}) can apply both to passive and active particles \cite{bechinger2016active}. To investigate the impact of social interaction, we use a passive model for simplicity, which allows us to apply Monte Carlo simulations.  We tested our model at different temperatures and density values. We found that the proposed model reproduces  radial distribution and angular correlation functions in qualitative agreement with empirical measurements (see SM Section D1). Moreover, for densities comparable to the observed empirical densities, we found  mono-peaked local density distributions for high-temperature values corresponding to a gas phase (Fig. 4a),  whereas bipeaked distributions for lower temperatures corresponding to a liquid-vapor coexistence phase (Fig. 4b). The insets of Fig. 4a-b demonstrates snapshots of movement patterns for these two phases.

\begin{figure}
  \includegraphics[width=1\linewidth]{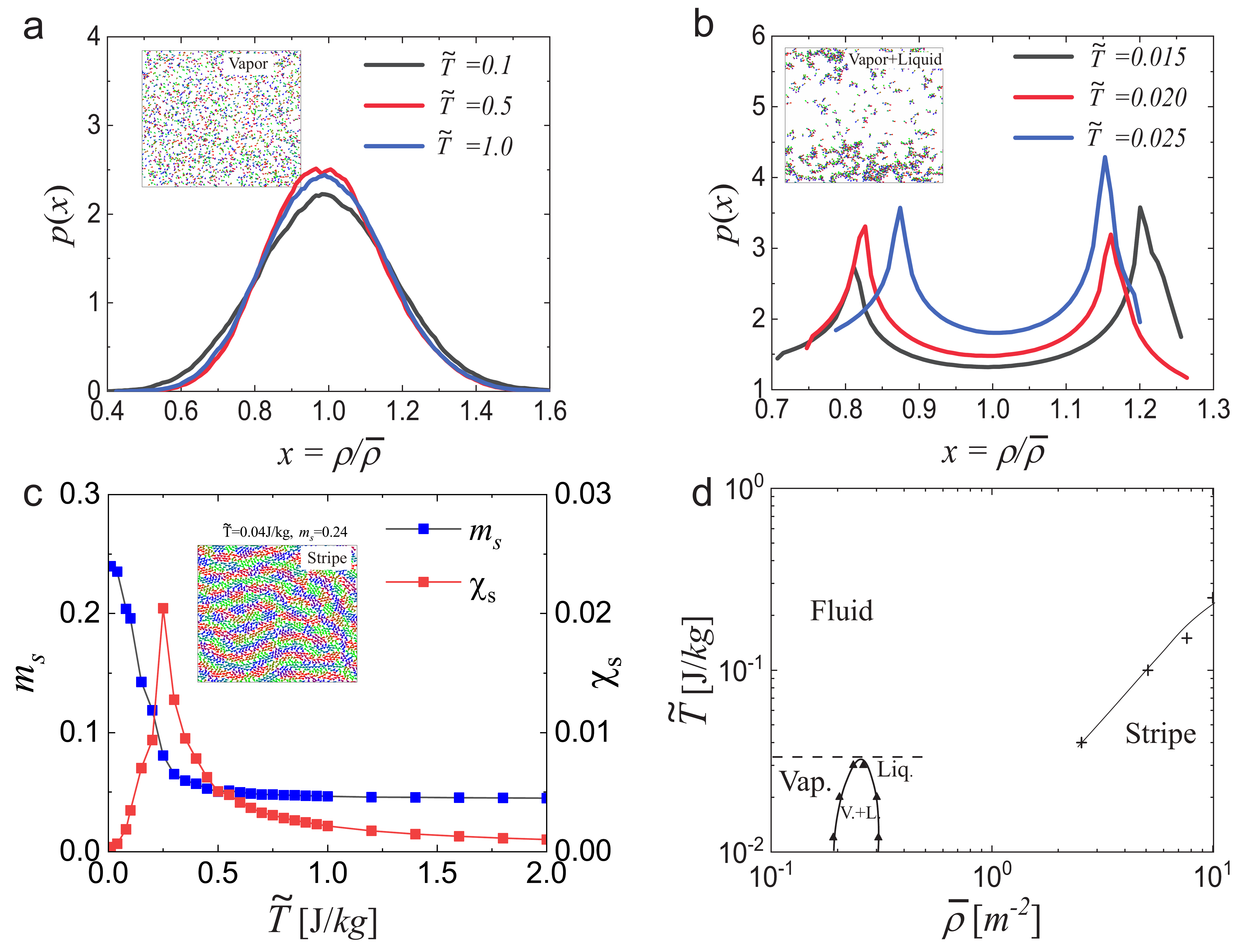}
  \caption{\textbf{Simulation results.} (a)-(b) show the local density distributions for relatively high and low temperatures with an average density ($0.25m^{-2}$), respectively. (c) Maximum mean-absolute order parameter $|M|_{max}$ as a function of effective temperature $\tilde{T}$ with a relatively high average density ($20m^{-2}$). (d) Density-temperature phase diagram. Scatter points are estimated by MC simulation. Black solid lines represent the boundaries among phases. Insets in (a)-(c) are the snapshots of corresponding systems.}
  \label{fig:simulations}
\end{figure}

%Old: Moreover, the classical $XY$ model predicts  the celebrated Berezinskii–Kosterlitz–Thouless transition between ordered and disordered magnetic phases \cite{kosterlitz1973ordering,jos201340}. While this phenomenon was not observed in both experiments and low density simulations, we do observe a similar phase transition occurs at high density values in our model. However,  unlike the XY model, the ordered phase of our model has a stripped pattern instead of a ferromagnetic one (see SM Section XX). This reflects the fact that our coupling function J(r) is ferromagnetic in short distances whereas antiferromagnetic in long distances.

%Old: Figure 4d summarizes the phase diagram for our model generated by numerical simulations. For a relative low density, the model reproduces the gas and vapor-liquid coexistence phases similar to those observed in experiments (see Fig. 3d). For a high density $\bar{\rho}\gg1$ which is beyond access of the existing datasets, the model predicts an ordered phase consisted of a striped pattern. Needless to say, this prediction requires further validation from future experiments.

Similar to the empirical dataset, there are no orientational order changes during the liquid-vapor phase transition in our model. It is interesting to explore such a magnetization phase transition, similar to the celebrated Berezinskii–Kosterlitz–Thouless transition observed in the classical $XY$-model \cite{stanley1971phase, jos201340} and existing social force models \cite{vicsek1995novel,helbing1995social,helbing2000simulating,helbing2001traffic}. Indeed, we observe a similar phase transition that occurs at very high-density values around $\bar{\rho}\gg1$, about one order of magnitude larger than that observed empirically. Nevertheless,  the magnetization order in our model shows a striped pattern instead of a ferromagnetic phase (see Inset of Fig. 4c). This reflects the fact that our coupling function $J(r)$ is ferromagnetic in short distances and antiferromagnetic at longer distances. Figure 4c plots the stripped order parameter $m_s$ and susceptibility $\chi_s$ as functions of temperature  (see SM Section D3 for definitions), providing clear evidence of such a phase transition. Note that this orientational phase transition occurs independently of the vapor-liquid phase transition, implying that the underlying physics of our model Eq. (\ref{eq:model}) is distinct from those of existing models \cite{vicsek1995novel,gregoire2004onset}.

Figure 4d summarizes the phase diagram for our model generated by numerical simulations. For a relatively low density, the model reproduces the gas and vapor-liquid coexistence phases similar to those observed in experiments (see Fig. 3d). For a high density $\bar{\rho}\gg1$ which is beyond the access of the existing datasets, the model predicts an ordered phase consisted of a striped pattern. Needless to say, this prediction requires further validation from future experiments.

In conclusion, we  utilized relatively high-resolution UWB-RFID technology to track human movements. We measured RDFs and angular correlation functions for two preschools and one kindergarten. Measurement of LDDs revealed two social phases, i.e., a gas-like phase corresponding to a single-peaked LDD and a liquid-vapor coexistence-like phase with a doubly peaked LDD. We measured the MSDs and VACFs which allowed us to calculate the effective temperature of classroom movement. Different social phases correspond to different effective temperatures and population densities, suggesting a phase diagram of human movements driven by social interaction. As noted by Ref~\cite{sinhuber2017phase}, the effective temperature measure described here could enrich our understanding of phase co-existence in insects and other social animals. 

Specifically, the liquid-vapor coexistence-like phase and the higher temperature gas-like system reflect differences in the observations of schools. The lower-temperature classrooms were acquired over a complete half-day observation  in which teachers interacted actively with children. This appeared to reduce effective temperature allowing  the formation of small dynamically stable social groups that are similar to liquid droplets. Free-moving individuals may join and leave these social groups with relatively high mobility, similar to vapor in the coexistence phase. Similar coexistence has been observed in swarm insects \cite{sinhuber2017phase}.  By contrast, the higher-temperature classrooms recorded the free-play activities of kindergarten children in which teachers were requested not to interact with children. Comparatively higher velocities and lower RDF peak values observed in these classrooms suggested a weaker social interaction, making the state of the system to be more gas-like.

In physical systems, individual properties change continuously and macroscopic systems show emergent changes in  their phases. Here we show that social systems may exhibit parallel properties. To the best of our knowledge, our work provides the first experimental evidence of the existence of multiple phases in human movements, a hypothesis posited theoretically two decades ago \cite{gregoire2004onset}. Future research is required to experimentally manipulate control parameters such as temperature that are hypothesized to control these phase transitions. 
Inspired by these empirical observations, we proposed a social interactive model that captures the nature of human orientational interactions at different distances. Our  model not only reproduces the observed social phases but also captures several physical quantities including radial distribution and angular correlation functions, in qualitative agreement with empirical measurements. The results shed light on the role of social interactions in human movement, with potential implementations in fields including  behavioral science \cite{saragosa2022real}, biological science and epidemiology \cite{zhang2022simulating}.

\begin{acknowledgments}
This work was supported partially by the National Science Foundation under Grants No.~2150830 and No.~IBSS-1620294, the Institute of Education Sciences under Grant No. ~R324A180203, and the National Institutes of Health under Grant No.~R01DC018542.
\end{acknowledgments}

\bibliography{ref}% Produces the bibliography via BibTeX.

\end{document}